\documentclass[%
prl%
 ,twocolumn%
 ,secnumarabic%
,tightenlines%
,amssymb, amsmath,nobibnotes, aps, prl, showpacs]{revtex4}

\usepackage{docs}%
\usepackage{bm}%
\usepackage{cf2}%
\usepackage{graphicx}

\def\la{\langle}

\def\la{\langle}

\def\degree#1{{^{\mathrm o} \mathrm#1}}
\expandafter\ifx\csname package@font\endcsname\relax\else
 \expandafter\expandafter
 \expandafter\usepackage
 \expandafter\expandafter
 \expandafter{\csname package@font\endcsname}%
\fi

\begin{document}


\title{Vorticity Alignment and Negative Normal Stresses in Sheared Attractive Emulsions}%

\author{Alberto Montesi$^{\ddag}$} 
\author{Alejandro A. Pe\~na$^{\dag}$}  
\author{Matteo Pasquali} \thanks{\texttt{mp@rice.edu}, $^{\dag}$\texttt{alejandr@rice.edu}, $^{\ddag}$\texttt{amontesi@rice.edu}} 
\affiliation{Department of Chemical Engineering - MS 362, Rice
University, 6100 Main St., Houston, TX 77005}
\date{\today}%

\begin{abstract}

Attractive emulsions near the colloidal glass transition are
investigated by rheometry and optical microscopy under shear. We
find that (a) the apparent viscosity $\eta$ drops with increasing
shear rate, then remains approximately constant in a range of
shear rates, then continues to decay; (b) the first normal stress
difference N$_1$ transitions sharply from nearly zero to negative
in the region of constant shear viscosity; (c) correspondingly,
cylindrical flocs form, align along the vorticity and undergo a
log-rolling movement. An analysis of the interplay between steric
constraints, attractive forces, and composition explains this
behavior, which seems universal to several other complex systems.


\end{abstract}

\maketitle

Emulsions are relatively stable dispersions of drops of a liquid
into another liquid in which the former is partially or totally
immiscible. Stability is conferred by other components, usually
surfactants or finely divided solids, which adsorb at the
liquid/liquid interfaces and retard coalescence and other
destabilizing mechanisms.

Emulsions can be regarded as repulsive or attractive, depending of
the prevailing interaction forces between drops.  In repulsive
emulsions, droplets repel each other at any center-to-center
distance. On the other hand, attractive emulsions exhibit a
potential well at a given distance that exceeds the energy
associated to random thermal fluctuations. Consequently, drops in
attractive emulsions form flocculates and gel-like structures,
whereas droplets in repulsive emulsions do not. This Letter
describes for the first time unusual rheological and
microstructural features in emulsions under shear, such as
alternating changes in the sign of the first normal stress
difference with increasing shear rate and formation of domains of
drops that align perpendicularly to the direction of shearing.  It
is shown that these features result from the interplay between
composition and attractive forces between droplets. Significant
similarities between these trends and those reported for different
systems, such as liquid crystalline polymers, colloidal
suspensions and polymeric emulsions, are also considered.

Experiments were performed on emulsions of bi-distilled water
dispersed in a lubricant oil base provided by Exxon Chemicals (
$\rho_{\textsf{oil}}$ = 871 kg/m$^3$, $\eta_{\textsf{oil}}$ =  91
mPa $\cdot$ s at 25 $\degree{C}$). The emulsions were stabilized
using the nonionic surfactant SPAN 80 (sorbitan monooleate, Sigma)
at a concentration of 5 wt.$\%$. Emulsification was carried out by
mixing in 1-inch internal diameter cylindrical plastic container
and blending with a two-blade paddle for 10 minutes at 1500 rpm.
Droplets in this emulsions form flocs mainly due to micellar
depletion attractions, because the concentration of surfactant is
well above the critical micellar concentration ($<$ 1 wt.$\%$).

Rheological measurements were carried out in a strain-controlled
ARES rheometer using several geometries \cite{rheometerfixtures}.
Microscopic observations were performed using a customized
rheo-optical cell consisting of two parallel glass surfaces, with
the upper one fixed to the microscope (Nikon Eclipse E600) and the
lower one set on a computer-controlled xyz translation stage
(Prior Proscan H101). The glass surfaces were rendered hydrophobic
to prevent adhesion and spreading of water drops. All tests were
performed at 25.0 $\pm$ 0.1 $\degree{C}$.

\begin{figure}
\includegraphics[height=5.9cm,width=7.4cm]{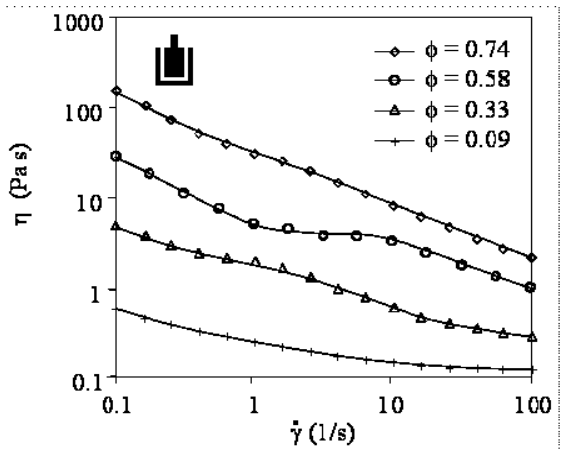}
\caption{Steady shear viscosity of water-in-oil attractive
emulsions formulated at several water contents. Measurements were
performed with the large concentric cylinder geometry.}
\label{different_conc}
\end{figure}

Figure \ref{different_conc} shows plots of shear viscosity $\eta$
vs. shear rate $\dot{\gamma}$ for four water-in-oil emulsions with
increasing volume fraction of the dispersed (water) phase $\phi$.
Emulsions for which $\phi$ = 0.09, 0.32 and 0.73 exhibited a
monotonic viscosity decay characteristic of flocculated emulsions
\cite{Pal1997}. The shear thinning is related to the progressive
breakdown of flocs at low and moderate $\dot{\gamma}$, and to the
deformation of drops at high $\dot{\gamma}$. The raise in $\eta$
with $\phi$ is a well-known effect caused by an increase in the
population of droplets, which leads to higher hydrodynamic
interactions among drops and higher resistance to flow. The
viscosity of the emulsion with $\phi= 0.58$ first decreased with
increasing shear rate, then remained relatively constant and then
continued to diminish. This trend has been reported for a few
attractive emulsions \cite{Pal1996,MokLee1997}, but it was not
explained. Here we explain the microstructural mechanism of the
transition in the trend of $\eta$, the corresponding anomalous
behavior of N$_1$ and the dependence of these phenomena on the
composition of the emulsion.

The rheological behavior of an emulsion made of two Newtonian
liquids shifts from viscous to predominantly elastic because of
changes in the arrangement of droplets from dilute (uncaged) to
caged, to packed, to compressed as the volume fraction of internal
phase $\phi$ grows \cite{Mason1999}. At the colloidal glass
transition ($\phi = \phi_\textsf{g} \sim 0.58$ \cite{Mason1997}),
droplets are caged indefinitely by their neighbors and random
thermal fluctuations do not disrupt such cages. The volume
fraction for close packing of monodisperse hard sphere
$\phi_{\textsf{cp}}$ ranges between 0.64 (random packing)
and 0.74 (ordered packing). 
When $\phi > \phi_{\textsf{cp}}$, the interfaces deform due to
compression of drops against drops. Figure \ref{different_conc}
shows that a region of constant $\eta$ occurs for attractive
emulsions in the transition from uncaged to compressed, with the
effect being more pronounced when $\phi \sim \phi_\textsf{g}$.

\begin{figure}
\includegraphics[height=5.85cm,width=7.4cm]{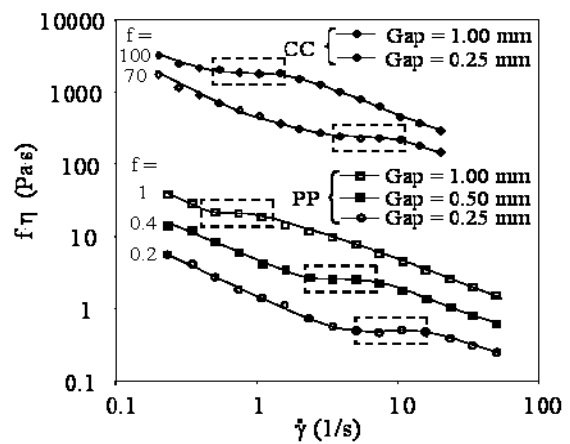}
\caption{Effect of the gap thickness on the shear viscosity
profile for emulsions with $\phi$ = 0.58 for two geometries. The
plots have been shifted by a factor f (indicated for each curve)
to facilitate interpretation.} \label{eta_diff_gap}
\end{figure}

Figure \ref{eta_diff_gap} shows the shear viscosity of emulsions
at $\phi$ = 0.58, measured with concentric cylinders (CC) and
parallel plates (PP) at several gap thicknesses. A region of
constant viscosity was observed in all cases. The range of shear
rates in which such phenomenon was observed shifted toward higher
$\dot{\gamma}$ as the gap size was reduced. Figure 2 also shows
that $\eta$ remained constant in a wider range of $\dot{\gamma}$
at narrower gap. Noticeably, the shear stress $\tau$ in the region
of constant $\eta$ always ranged between 6 and 30 Pa,
independently of geometry and gap size. This fact suggests that
the onset of a plateau in $\eta$ is related to changes in the
microstructure of the emulsion and that the structural transition
involves the formation of domains with characteristic size
influenced by the
gap size. 

\begin{figure}
\includegraphics[height=6.4cm,width=7.8cm]{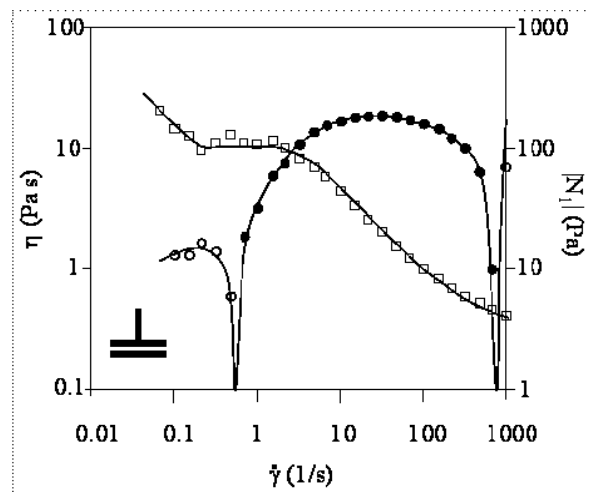}
\caption{Steady shear viscosity (squares) and magnitude of the
first normal stress difference (circles) vs. shear rate trends for
an emulsion with $\phi$ = 0.58. Filled circles correspond to
negative N$_1$. Measurements were performed with parallel-plate
geometry (gap = 0.65 mm).} \label{eta_and_N1}
\end{figure}

Figure \ref{eta_and_N1} shows $\eta$ and N$_1$ as a function of
the effective $\dot{\gamma}$ between parallel plates for an
emulsion with $\phi$ = 0.58. N$_1$ evolved from nearly zero to
negative, then from negative to positive as $\dot{\gamma}$ was
increased; the sharp transition to negative values of N$_1$
occurred concomitantly with the plateau in $\eta$ at each gap
size. Experiments with the cone-and-plate geometry, where
$\dot{\gamma}$ is uniform across the sample, showed the same
effects. We investigated the minimum value of negative normal
stresses N$_{1,min}$ as a function of the thickness of the gap
between the parallel plates. N$_{1,min}$ was always negative and
above the sensitivity of the instrument, and its absolute value
grew from 50 to 316 Pa as the gap was reduced from 1.0 to 0.5 mm.

\begin{table}
\begin{center}
\centering
\begin{tabular}{|c|c|c|c|c|c|c|}
\hline 
  \textbf{$\phi$}& $0.32$ & $0.52$ & $0.58$ & $0.62$ & $0.73$ & $0.75$\\
\hline 
  \textbf{N}$_{1,min}$ \textbf{(Pa)} & $ -24$ & $-238$ & $-210$ &  $-223$ & $-159$ & $ -50$\\
\hline
\end{tabular}
\end{center}
\caption{Minimum N$_1$ vs. water volume fraction $\phi$ for
pre-sheared emulsions (gap = 0.6 mm). Error in N$_{1,min}$
measurements = $\pm$ 30 Pa.} \label{N1_table}
\end{table}

The same trends in N$_1$ reported in Figure \ref{eta_and_N1} were
also observed in tests on emulsions with $\phi = 0.52$ and $0.62$,
with comparable minimum values of N$_1$, as shown in Table
\ref{N1_table}. At $\phi = 0.73$ and $0.75$ the transition to
negative N$_1$ was much less pronounced and observed in a much
narrower region of effective $\dot{\gamma}$. Emulsions with $\phi
= 0.32$ did not exhibit any of these features, and N$_1 \simeq 0$
at all $\dot{\gamma}$. These results indicate that the onset of
negative N$_1$ is favored by the caged configuration that droplets
adopt near the glass transition. The data reported in Figure
\ref{eta_and_N1} and Table \ref{N1_table} were obtained after
pre-shearing the emulsion for 60 s at $\dot{\gamma}$ =100
s$^{-1}$. This protocol was adopted to improve the reproducibility
of N$_1$ measurements. We observed similar trends for N$_1$ in
emulsions that were not pre-sheared, but the reproducibility of
N$_1$ values was poor. We have also formulated and tested
repulsive emulsions that matched the composition, the viscosity
ratio of the phases and the interfacial tension of the attractive
emulsions studied here. The effects described above were not
observed in such repulsive emulsions, thus confirming the key role
of attractive forces.

\begin{figure}
\includegraphics[height=12.3cm,width=7.3cm]{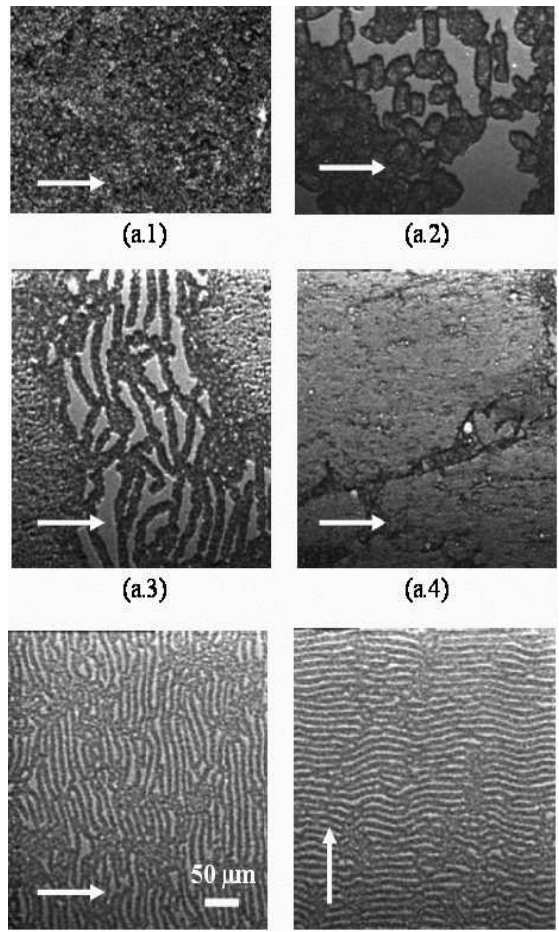}
\caption{Microphotographs of emulsions with  $\phi$ = 0.58 under
simple shear for gap thicknesses of (a) 20 $\mu$m; (b) 12 $\mu$m.
Arrows indicate the direction of shear. The reference scale for
all frames is inserted in frame (b.1.).} \label{microstructure}
\end{figure}

Figure \ref{microstructure} summarizes microscopic observations on
the arrangement of droplets under simple shear for an attractive
emulsion with $\phi = 0.58 $ using the rheo-optical cell described
above. Frames (a.1) through (a.4) were taken at increasing shear
rates ($\sim$ 0.5, 5, 80 and 120 s$^{-1}$, respectively). At low
$\dot{\gamma}$ (frame (a.1)), the system exhibited a fairly
uniform arrangement of droplets. Voids filled with continuous
phase (oil) between flocs were present sparsely. When
$\dot{\gamma}$ was raised, large flocs disaggregated to form
domains surrounded by an increasing number of voids. Speckle-like
flocs detached from such domains and tumbled as the emulsion was
sheared (frame a.2). At higher $\dot{\gamma}$, the speckles
aligned 
along the direction of vorticity, to form cylindrical flocs that
exhibited a ``rolling'' motion (frame a.3). These cylindrical
flocs were still surrounded by large uniform domains. At higher
shear rates, the structured configuration of flocs was disrupted
(frame a.4) and a nearly uniform arrangement of drops was
restored. Frames (b.1) and (b.2) illustrate that the formation of
banded flocs in the direction of vorticity was observed
independently of the direction in which the emulsion was sheared.
In these cases the ordered structure was observed throughout the
entire domain. The latter frames were taken with a gap thickness
narrower than the one imposed on frame (a.3). The comparison
between frames (a.3) and (b.1) suggests that a reduction in the
gap led to the formation of finer, more ordered and more evenly
distributed band structures. Also, the formation of large voids
filled with oil at intermediate shear rates suggests that droplets
within the flocs adopted a close-packed configuration as the
transition took place. Voids were filled with the oil that came
out of the flocs as the local concentration of droplets increased
within them. These voids disappeared at high $\dot{\gamma}$
because the shearing disrupted
the flocs. 

\begin{figure}
\includegraphics[height=6.1cm,width=8.4cm]{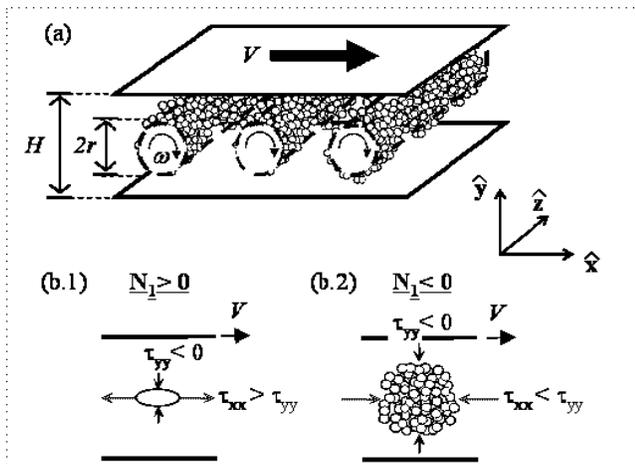}
\caption{Schematic representation of (a) the conformation of
cylindrical flocs and (b) the stresses in the plane of shear}
\label{schematic}
\end{figure}

The critical shear rate $\dot{\gamma}_{\textsf{C}}$ at which the
disruption of the ordered structure takes place can be estimated
from continuity of shear stress at the outer interface between the
flocs and the oil, assuming that the flocs undergo rigid-body
rotation with angular velocity $\omega$ (Fig. \ref{schematic}.a):
\beqa \label{stress_continuity} \nonumber \tau =
\tau_{xy,\textsf{oil}} &=& \tau_{xy, \textsf{floc}}  \\  =
\eta_{\textsf{oil}} \dot{\gamma}_{\textsf{eff}}
 &=& \eta_{\textsf{oil}} \left[ \frac{\dot{\gamma} - \left( 2r/H \right)
\omega}{1 - \left( 2r/H \right)} \right] \eeqa where $\tau$ is the
shear stress and $\dot{\gamma}_{\textsf{eff}}$ is the effective
shear rate. When $\dot{\gamma} = \dot{\gamma}_{\textsf{C}}$,
$\tau$ matches the yield stress of the flocs at the close-packing
composition, $\tau^{\textsf{cp}}_{\textsf{yield}}$, and $0 \le
\omega \le \dot{\gamma}_{\textsf{C}}$. Therefore, from Eq.
\ref{stress_continuity} we obtain: \beq \label{gamma_estimate}
\dot{\gamma}_{\textsf{eff}} \left[ 1 - (2r/H) \right] \leq
\dot{\gamma}_{\textsf{C}} \leq \dot{\gamma}_{\textsf{eff}};
\dot{\gamma}_{\textsf{eff}} =
\frac{\tau^{\textsf{cp}}_{\textsf{yield}}}{\eta_{\textsf{oil}}}.
\eeq

Eq. \ref{gamma_estimate} can be validated with the experiment
shown in Fig. \ref{microstructure}, frames (a.1-a.4). In this test
the cylindrical flocs were disaggregated near  $\dot{\gamma} \sim
95$ s$^{-1}$. From the frame (a.3) in Fig. \ref{microstructure} we
obtain $2r/H \sim 0.6$. Also, we measured
$\tau^{\textsf{cp}}_{\textsf{yield}} \sim 15$ Pa for
$\phi_{\textsf{cp}} \sim 0.74$. With these figures and
$\eta_{\textsf{oil}}$ = 0.091 Pa $\cdot$ s, Eq.
\ref{gamma_estimate} renders 68 s$^{-1} <
\dot{\gamma}_{\textsf{C}} < 165$ s$^{-1}$.

The features described above exhibit striking similarities with
the rheological properties of nematic liquid-crystalline polymers
(LCP), namely \cite{Marrucci1991,Kiss1998,Larson1999}: (a) $\eta$
vs. $\dot{\gamma}$ plots for LCP solutions commonly exhibit two
shear-thinning regions with an intermediate zone of constant
viscosity;  (b) changes in sign of N$_1$ first positive to
negative, and then negative to positive as $\dot{\gamma}$ is
increased are observed for some of these systems; (c) tumbling
domains and formation of bands normal to the plane of shear are
reported for several lyotropic and thermotropic LCP under shear,
sometimes in connection with the above-mentioned trends. Domains
of particles that extend along vorticity under shear have also
been reported for weakly flocculated magnetic suspensions
\cite{Navarrete1992} and thixotropic clay gels \cite{Pignon1997}.
In these latter cases, as well as in the emulsions tested here,
attractive forces between particles determined the formation of
such domains.

The complex rheological behavior of attractive emulsions reported
in this Letter is due to several concomitant factors. First,
attractive forces cause the formation of flocs and confer them
elasticity. Second, steric constraints impose diameter and
ordering of flocs. Third, normal stresses in the shearing plane
arise due to the flocs elasticity and make them extend along the
vorticity. Elongation of drops along the vorticity was observed in
polymeric emulsions \cite{Hobbie1999}, and was related to higher
normal forces in the dispersed phase relative to the continuous
one, which compressed the drops in the shearing plane and caused
transverse elongation, in a fashion similar to the phenomenon of
rod climbing. The effect is also observed here, with the flocs
acting as the viscoelastic dispersed phase. Compression in the
$\hat{x}$-$\hat{y}$ shearing plane (Fig. 5.b.2.) leads to a
close-packed configuration within the flocs, and to segregation of
continuous phase to the surrounding voids. The effect is neither
observed in uncaged emulsions because the average distance between
drops is large and the flocs are not forced sterically to form
compact structures, nor in highly concentrated emulsions because
the compressed arrangement of drops inhibits the formation of
voids and thus the structural transition. The onset of negative
N$_1$ seems intrinsic to the conformation of flocs along the
vorticity. A single drop under simple shear (Figure 5.b.1) is
subject to compression in the $\hat{y}$ direction ($\tau_{yy} <
0$) and tension in the $\hat{x}$ direction ($\tau_{xx} > 0$)
\cite{Schowalter1978}. Consequently, the drop elongates in the
direction of shearing and N$_1 = \tau_{xx} - \tau_{yy} > 0$. On
the other hand, the floc is subject to compression in both
directions as discussed above (Figure 5.b.2), so  $\tau_{xx} < 0$
and $\tau_{yy} < 0$. The growth of the flocs along $\hat{y}$ is
restricted by geometry, and growth along $\hat{x}$ is prevented if
$\tau_{xx} < \tau_{yy}$ (i.e., N$_1 < 0$). Thus, the flocs can
grow in size only along the neutral direction $\hat{z}$. This
explains the onset of negative N$_1$ at the intermediate
$\dot{\gamma}$ at which cylindrical flocs form.

The authors wish to acknowledge E. K. Hobbie, K. B. Migler, T. G.
Mason, and C. W. Macosko for useful discussions. This work was
partially supported by the National Science Foundation
(CTS-CAREER-0134389), the Lodieska Stockbrigde Vaughan Fellowship
(AAP) and the Rice University Consortium for Processes in Porous
Media (G. J. Hirasaki, director).

\bibliography{amontesi}

\end{document}